\documentstyle[eqsecnum,aps]{revtex}  

\def\lsim{\mathrel{\hbox{\rlap{\hbox{\lower4pt\hbox{$\sim$}}}\hbox{$<$}}}}
\def\gsim{\mathrel{\hbox{\rlap{\hbox{\lower4pt\hbox{$\sim$}}}\hbox{$>$}}}}

\begin{document}

\renewcommand{\thefootnote}{\fnsymbol{footnote}}

\title{\bf Constraints on the chaotic inflationary scenario with a nonminimally
coupled ``inflaton'' field from the cosmic microwave background radiation anisotropy}

\author{Eiichiro Komatsu\footnotemark[1]  and    
Toshifumi Futamase\footnotemark[2]}

\address{\small\sl Astronomical Institute, Graduate School of Science, 
Tohoku University, Sendai 980-77, Japan}                      
\footnotetext[1]{E-mail: komatsu@astr.tohoku.ac.jp}
\footnotetext[2]{E-mail: tof@astr.tohoku.ac.jp}

\maketitle

\begin{abstract}

We investigate a possibility to restrict the chaotic inflationary 
scenario with large nonminimally coupled inflaton field $\phi$ 
considered by Fakir and Unruh by means of the observed cosmic microwave background radiation(CMBR) anisotropy. This model is characterized by the condition $\xi>1$ and $\psi\equiv 8\pi G\xi\phi^2 \gg 1$ where $\xi$ is the nonminimal coupling constant. We calculate the contributions 
of the long wavelength gravitational waves (GW) generated in the 
inflationary period to CMBR anisotropy quadrupole moment. 
We obtain the constraint  $\lambda/\xi^2 =
1.8\times 10^{-8}/\psi_{i}$, where $\lambda$ is the self coupling and 
$\psi_i$ means the initial value of $\psi$. Combining this 
with previously obtained constraint $\sqrt{\lambda/\xi^2} \approx (\delta T/T)_{\rm rms} = 1.1\times 10^{-5}$, we conclude that the initial value 
has to be  $\psi_i \approx 1.6\times 10^3$. 
If the self-coupling has a reasonably values of order $10^{-2}$, then 
$\xi \approx 10^4$ and $\phi_i \approx 10^{-1} m_{pl}$ where $m_{pl}$ is 
the Planck mass.

\pacs{PACS number(s): 04.50.+h, 98.80.-k, 98.70.Vc}

\end{abstract}

\section{Introduction}
\indent
Inflation is regarded as one of the most important conceptual progress  
in modern cosmology\cite{Sato,Guth}. 
This concept give us not only the solution of cosmological puzzles 
such as the horizon and flatness problem, but also the origin of the cosmological perturbations. 
Among various scenarios of inflation proposed so far, 
Linde's chaotic scenario seems to be a natural mechanism for the realization 
of the inflationary expansion\cite{Linde83}. 
In spite of many attractive features, the original scenario still have the fine tuning problem of the parameters. Namely one has to choose 
an unreasonably small self coupling constant, i.e. $\lambda < 10^{-12}$, 
to have the correct magnitude of the density perturbation which is 
consistent with the observed tiny CMBR fluctuations.

In order to overcome the difficulty, 
Fakir and Unruh proposed a new version of the chaotic model 
which has {\sl strong} nonminimal coupling, i.e. $\xi>1$, to the scalar
curvature\cite{FU90}. 
Note that we shall adopt in the  present paper the sign convention 
for $\xi$ such that the conformal coupling means $\xi = -1/6$. The chaotic inflationary scenario with nonminimal coupling has been investigated for 
various values of $\xi$ by Futamase and Maeda\cite{FM89}. Their 
result does not exclude the {\sl strong} coupling model of Fakir and Unruh. 
They have shown  that the larger value of $\xi$ relaxes the 
fine-tuning problem of the self-coupling $\lambda$, and this
conclusion has been followed by Makino and Sasaki more
rigorously\cite{MS91}. 

Here we shall investigate Fakir-Unruh scenario from different point of view. Namely we shall study the effect of the scenario on the CMBR anisotropy. 
It is know that the long wavelength gravitational waves are generated during 
the period of the inflationary expansion, and they make 
a significant contribution to  CMBR anisotropy through the Sachs-Wolfe effect\cite{SW67} on lower multipoles\cite{AW84,S}. 
We will point out that Fakir-Unruh scenario  has also a nonnegligible contribution to the spectrum of CMBR anisotropy by generated GW's, 
and then obtain the constraint on a certain combination of parameters 
by comparing the predicted quadrupole moment with the results of COBE-DMR observations.

This paper is organized as follows. We first review Fakir-Unruh scenario 
and present the self-consistent inflationary solutions in Section 2. 
We then find the radiative modes of the metric perturbations in the spacetime generated by nonminimally coupled inflaton field, and  
derive the resulting spectrum of GW in section 3.  Based on the result 
in section 3, we shall calculate the power spectrum of CMBR anisotropy 
and compare it with the observations of COBE-DMR in Section 4.  
Finally some discussions are made in section 5. We shall follow 
Misner, Thorne, and Wheeler\cite{MTW} for the definition of 
Riemann tensor, Ricci tensor and Ricci scalar.

\section{Background Inflationary Solutions}

As explained in the introduction, we shall be in this paper interested in 
the chaotic inflationary scenario with large positive nonminimal coupling. 
Thus we shall take the  following action as our model.
\begin{equation}\label{action}
S = \int d^4x\sqrt{-g}\left[\frac R{2\kappa^2}+\frac12\xi\phi^2R-\frac12\phi_{,\mu}\phi^{,\mu}+V(\phi)\right],
\end{equation}
where $\kappa^2\equiv 8\pi G$. 
Our definition of $\xi$ is the same as Fakir and Unruh, that is,
conformal coupling means $\xi = -\frac16$. 
Note that Futamase and Maeda used opposite sign for $\xi$. 
Varying the action (\ref{action}), we obtain the field equations
\begin{eqnarray}
\left(\frac1{\kappa^2}+\xi\phi^2\right)G_{\mu\nu} 
& = & (1+2\xi)\phi_{,\mu}\phi_{,\nu}
      -\left(2\xi+\frac12\right)
             g_{\mu\nu}\phi_{,\alpha}\phi^{,\alpha} \nonumber \\
  \mbox{} & - & 2\xi\phi\left(g_{\mu\nu}\Box
                 -\nabla_{\mu}\nabla_{\nu}\right)\phi 
      +g_{\mu\nu}V(\phi)\label{Eins}, \\
\Box\phi+\xi R\phi+V_{,\phi} 
& = & 0\label{scalar},
\end{eqnarray}
where $\Box \equiv {-g}^{-1/2} \partial_{\alpha}\left({-g}^{1/2}g^{\alpha\beta}\partial_{\beta}\right)$, 
and $V_{,\phi} = {\partial V}/{\partial \phi}$. Greek and Latin
indices take $0,1,2,3$ and $1,2,3$, respectively.

For the spacetime we assume a homogeneous and spatially 
flat Robertson-Walker universe
\begin{equation}
{\rm d}s^2 = {\rm d}t^2-a^2(t)\delta_{ij}{\rm d}x^i{\rm d}x^j.
\end{equation}
Equation(\ref{Eins}) yields the Hamiltonian constraint equation
\begin{equation}
H^2=\frac{\kappa^2}{3(1+\kappa^2\xi\phi^2)}\left[\frac12{\dot{\phi}}^2+V(\phi)-6\xi H\phi\dot{\phi}\right],
\end{equation}
where overdots denote time derivatives.  
The momentum constraint equation is
\begin{equation}
\dot{H}=\frac{\kappa^2}{1+\kappa^2\xi\phi^2}\left[-\frac12{\dot{\phi}}^2+\xi H\phi\dot{\phi}-\xi{\dot{\phi}}^2-\xi\phi\ddot{\phi}\right],
\end{equation}
The equation of motion for the inflaton field
is obtained from (\ref{scalar}) as 
\begin{equation}
\ddot{\phi}+3H\dot{\phi}-6\xi\left(\dot{H}+2H^2\right)\phi+V_{,\phi}=0.
\end{equation}
Combining above equations, we obtain
\begin{eqnarray}
\ddot{\phi}+3H\dot{\phi}
& + & \left[\frac{\kappa^2\xi\phi^2(1+6\xi)}
 {1+\kappa^2\xi\phi^2(1+6\xi)}\right]\frac{\dot{\phi}^2}{\phi} \nonumber \\
\mbox{} &=&\frac1{1+\kappa^2\xi\phi^2(1+6\xi)}\left[4\kappa^2\xi\phi V(\phi)-(1+\kappa^2\xi\phi^2)V_{,\phi}\right].
\end{eqnarray}
Using the "slow-roll approximations" such as 
\begin{eqnarray}
\left|\frac{\ddot{\phi}}{\dot{\phi}}\right|&\ll& H,\\
\left|\frac{\dot{\phi}}{\phi}\right|&\ll& H,\\
\frac12{\dot{\phi}}^2&\ll& V(\phi),
\end{eqnarray}
the energy constraint and field equation take the following forms, respectively
\begin{eqnarray}\label{H^2}
H^2&\approx& \frac{\kappa^2}{3(1+\kappa^2\xi\phi^2)} \nonumber \\
\mbox{} & & \left[V(\phi)-\frac{2\xi\phi}
     {1+\kappa^2\xi\phi^2(1+6\xi)}\left\{4\kappa^2\xi\phi         
          V(\phi)-(1+\kappa^2\xi\phi^2)V_{,\phi}\right\}\right]\\
3H\dot{\phi}&\approx& \frac1{1+\kappa^2\xi\phi^2(1+6\xi)}\left[4\kappa^2\xi\phi V(\phi)-(1+\kappa^2\xi\phi^2)V_{,\phi}\right]\label{phi},
\end{eqnarray}
These are the basic equations which determine the background solution.
We shall consider the solutions of these equations in the case with the  self-coupling constant and mass term, separately.

\subsection{Chaotic Inflation by a self coupling}

Let us consider the chaotic inflation generated by the self-coupling constant 
of the inflaton. Thus we take
$$V(\phi)=\frac14\lambda\phi^4.$$
Then the equations (\ref{H^2}) and (\ref{phi}) give, respectively
\begin{eqnarray}
H^2 &\approx& \frac{\kappa^2\lambda\phi^4} 
 {12(1+\kappa^2\xi\phi^2)}\left[1+\frac{8\xi}
{1+\kappa^2\xi\phi^2(1+6\xi)}\right],\\
3H\dot{\phi}&\approx& -\frac{\lambda\phi^3}{1+\kappa^2\xi\phi^2(1+6\xi)},
\end{eqnarray}
Let us now define our basic variable $\psi\equiv \kappa^2\xi\phi^2$ 
which turns out to be very useful in our analysis. Since we consider
the {\sl strong} curvature coupling, let us consider the situation where 
$$\psi \gg 1.$$
This condition simplifies significantly the above equations as
\begin{eqnarray}
H^2 &\approx& \frac{\lambda\psi}{12\kappa^2\xi^2},\label{expansion}\\
\frac{\dot{\psi}}{\psi}&\approx& -\frac{2\lambda}{3\kappa^2\xi(1+6\xi)}\frac1H,
\end{eqnarray}
Therefore we have the following solution.
\begin{eqnarray}
\frac{\dot{\psi}}H&\approx& -\frac{8\xi}{1+6\xi},\\
\psi&\approx& \psi_i-\frac{8\xi}{1+6\xi}\ln{\left(\frac a{a_i}\right)}\label{psi}.
\end{eqnarray}
Note that this solution is consistent with the slow-roll approximations. 
In fact we have
\begin{equation}
\left|\frac{\dot{\psi}}{H\psi}\right|\approx \left|-\frac{8\xi}{1+6\xi}\frac1{\psi}\right| \ll 1.
\end{equation}
From the equations (\ref{expansion}) and (\ref{psi}), 
we obtain a differential equation for the scale factor.
\begin{equation}\label{differ}
\frac  {\dot{a}}a=\sqrt{\frac{\lambda\psi_i}{12\kappa^2\xi^2}}\left[1-\frac{8\xi}{(1+6\xi)\psi_i}\ln{\left(\frac a{a_i}\right)}\right]^{1/2}.
\end{equation}
This equation (\ref{differ}) can be easily integrated as
\begin{equation}
a(t)=a_i\exp{\left[H_i t-\gamma(H_it)^2\right]},
\end{equation}
Thus we have the following  Hubble parameter and scalar field.
\begin{eqnarray}
a(t)&=&a_i\exp{\left[H_i t-\gamma(H_it)^2\right]},\\
H(t)&=&H_i\left(1-2\gamma H_it\right),\\
\psi(t)&=&\psi_i\left(1-2\gamma H_it\right)^2,
\end{eqnarray}
where
\begin{eqnarray}
H_i^2&=& \frac{\lambda\psi_i}{12\kappa^2\xi^2},\\
\gamma&=& \frac{2\xi}{(1+6\xi)\psi_i}.
\end{eqnarray}
These expressions indicate that the expansion rate is much more rapid 
than the rate of change in the Hubble parameter as well as  the inflaton 
field. Since we are interested in the very long-wavelength tensor perturbations 
which give the largest contribution to the CMBR anisotropy today and 
such perturbations decouple at the early stage of inflation, 
we can approximate the above solution to the following simple forms:
\begin{eqnarray}
a(t)&\approx &a_ie^{H_i t},\\
H(t)&\approx &H_i, \\
\psi(t)&\approx &\psi_i \label{psiconst}.
\end{eqnarray}

Finally, we have to consider the amount of inflation sufficient to solve cosmological puzzles,
\begin{eqnarray}
\frac{a(t)}{a_i}&=&\exp\left(\int Hdt\right)=\exp\left(\int_{\psi_i}^{\psi_f}\frac{H}{\dot{\psi}}d\psi\right)\nonumber\\
&\approx& \exp\left(\int_{\psi_i}^{\psi_f} -\frac{1+6\xi}{8\xi}d\psi\right)=\frac{1+6\xi}{8\xi}(\psi_i-\psi_f) \geq 70.
\end{eqnarray}
Fakir and Unruh commented $\psi \approx 1$ signals the end of inflation\cite{FU90}. 
Thus the {\sl strong} coupling, i.e. $\xi>1$, condition gives
\begin{equation}\label{infamount}
\psi_i \gsim 80.
\end{equation}

\subsection{Chaotic Inflation by a mass term}

Next, we consider a massive free inflaton field with the following potential
$$V(\phi)=\frac12m^2\phi^2$$ 
Then the equations (\ref{H^2})and (\ref{phi}) become, respectively
\begin{eqnarray}
H^2&\approx& \frac{\kappa^2\xi m^2\phi^2}{6\xi(1+\kappa^2\xi\phi^2)}\left[1+\frac{4\xi}{1+\kappa^2\xi\phi^2(1+6\xi)}(1-\kappa^2\xi\phi^2)\right],\\
3H\dot{\phi}&\approx& \frac{m^2\phi(\kappa^2\xi\phi^2-1)}{1+\kappa^2\xi\phi^2(1+6\xi)}.
\end{eqnarray}
When $\psi\gg 1$, these equations may be approximated as 
\begin{eqnarray}
H^2&\approx& \frac{m^2}{6\xi}\left(\frac{1+2\xi}{1+6\xi}\right),\\
\frac{\dot{\psi}}{\psi}&\approx&\frac{2m^2}{3(1+6\xi)}\frac1H.
\end{eqnarray}
We can find a scalar field grows exponentially
so that the slow-roll approximations are not satisfied
unless $\xi\ll 1$ :
\begin{equation}
\left|\frac{\dot{\psi}}{H\psi}\right|\approx \left|\frac{4\xi}{1+2\xi}\right| 
< 2.
\end{equation}
Thus, we conclude the mass-term inflation does not occur
in the model with {\sl strong} coupling. 
This result is consistent with Futamase and Maeda\cite{FM89}.

\section{The Spectrum of GWs generated during the Inflationary Expansion}

In this section we shall calculate the spectrum of GWs generated during the period od inflationary expansion 
in the framework of Fakir-Unruh scenario.

\subsection{Radiative Solutions of Metric Perturbations}

First we derive the equation for the GWs as the linear metric perturbation. 
The method is exactly same with the case in general relativity(GR)\cite{Wein}. 
Namely we consider a small disturbance $h_{\mu\nu}$ of the spatially flat Friedmann-Robertson-Walker metric $\bar{g}_{\mu\nu}$:
\begin{equation}
ds^2 = (\bar{g}_{\mu\nu}+h_{\mu\nu})dx^{\mu}dx^{\nu}.
\end{equation}
The spatial flatness means that the metric perturbations can be expanded by a plane wave with a comoving wave number $k$ :
\begin{equation}
h^\mu_\nu(\mbox{\boldmath$x$},\tau)=\sum_\lambda \int{{\rm d^3}\mbox{\boldmath$k$} h_{\lambda}(\mbox{\boldmath$k$},\tau)e^{i\mbox{\boldmath$k$}\cdot\mbox{\boldmath$x$}}\epsilon^\mu_\nu},
\end{equation}
where $\epsilon^\mu_\nu$ is the polarization tensor, 
and $\lambda=+,\times $ is a mode of the polarization. 
We choose the synchronous gauge ($h_{00}=h_{0i}=0$), and use the
conformal time ($x^0 = \tau$), 
\begin{equation}
{\rm d}s^2 = a^2(\tau){\rm d}\tau^2-\left(a^2(\tau)\delta_{ij}+h_{ij}\right){\rm d}x^i{\rm d}x^j.
\end{equation}
Then the linearized Einstein equations (\ref{Eins}) lead to the radiative mode of 
the tensor perturbation: 
\begin{equation}
h^k_k=h^k_{i,k}=\delta\psi=0{\rm ,}
\end{equation}
\begin{equation}\label{maineq}
h_{\lambda}''(\mbox{\boldmath$k$},\tau)+\left(2\frac{a'}a
+\frac{\psi'}{1+\psi}\right)h_{\lambda}'(\mbox{\boldmath$k$},\tau)
+k^2h_{\lambda}(\mbox{\boldmath$k$},\tau)=0,
\end{equation}
where dashes denote conformal time derivatives. 
The other components of the Einstein equations (\ref{Eins}) are trivially satisfied with above conditions.

Writing each component $h_{\lambda}(\mbox{\boldmath$k$},\tau)$ of the
GW perturbations as \cite{GG93,BMM93},
\begin{equation}
h_{\lambda}(\mbox{\boldmath$k$},\tau)\equiv\frac1{R(\tau)}\mu_{\lambda}(\mbox{\boldmath$k$},\tau)=
\frac1{a\sqrt{1+\psi}}\mu_{\lambda}(\mbox{\boldmath$k$},\tau),
\end{equation}
we have the following equation for $\mu_\lambda$.
\begin{equation}\label{Sch}
\mu_{\lambda}''(\mbox{\boldmath$k$},\tau)
+\left(k^2-\frac{R''}R\right)\mu_{\lambda}(\mbox{\boldmath$k$},\tau)=0. 
\end{equation}
This is an ordinally "Schr\"odinger type" equation. If $k^2\gg
|{R}''/R|$, we can find a plane wave solution of
$h_{\lambda}$ with its amplitude scaled by 
$\left(a\sqrt{1+\psi}\right)^{-1}$.

\subsection{The Spectrum of GWs}

Now we will calculate the spectrum of the GWs derived above. 
We follow the previous works\cite{AW84,AH86,White92}. 
Let us choose the normalization of $a(\tau)$ as 
\begin{eqnarray}
a(\tau)&=&
-\frac1{H_i\tau}\qquad{\rm vacuum}\qquad\tau\in (-\infty,-\tau_2),\\
&= &\frac{2\tau_1\tau}{\tau_0^2}\qquad {\rm radiation}\qquad\tau\in (\tau_2,\frac{\tau_1}2),\\
&= &\frac{\tau^2}{\tau_0^2}\qquad {\rm matter}\qquad\tau\in (\tau_1,\tau_0).
\end{eqnarray} 
We assumed that the transitions across each phase are instantaneous 
and we matched $a(\tau)$ and $\dot{a}(\tau)$ at the transition
points. This will be an accurate approximation for the long-wave tensor perturbations we are interested in. 
Note that $\tau$ is discontinuous across each transitions. 

We shall calculate the amplitude of the GWs generated 
as quantum noises. The following consideration will help us to 
choose a convenient field variable to quantize. 
The equation (\ref{Sch}) shows that each polarization state of the
waves behave as a scalar field $\varphi$ with a
normalization factor of $\sqrt{2 \kappa^2_{\rm eff}}$:
\begin{equation}\label{norm}
h_{\lambda}=\sqrt{2 \kappa^2_{\rm eff}}\varphi_{\lambda}
=\sqrt{\frac{2 \kappa^2}{1+\psi}}\varphi_{\lambda}.
\end{equation}
We omit a subscript $\lambda$ henceforth. 
This situation is analogous to GR case\cite{Grishchuk}. 
We also neglected the other interactions of $\varphi$ due to the mixture of 
$\psi'$, which is a good approximation for our purposes. 
Thus we choose the scalar field $\varphi$ to quantize as the same way in GR.  
\begin{equation}
\varphi(\mbox{\boldmath$x$},\tau)
=\int{{\rm d^3}\mbox{\boldmath$k$}
  \left[ \hat{a}_{\rm phase}(\mbox{\boldmath$k$})\varphi_k(\tau)
   e^{i\mbox{\boldmath$k$}\cdot\mbox{\boldmath$x$}}
    + \hat{a}^{\dag}_{\rm phase}(\mbox{\boldmath$k$})\varphi^*_k(\tau)
      e^{-i\mbox{\boldmath$k$}\cdot\mbox{\boldmath$x$}}\right]},
\end{equation}
where $\hat{a}_{\rm phase}$ and $\hat{a}^{\dag}_{\rm phase}$ are 
the annihilation and  creation operator 
at each phase(vacuum dominated, radiation dominated and 
matter dominated phases), and $\varphi_k(\tau)$ is properly normalized solution of the equation (\ref{maineq}) :
\begin{eqnarray}
\varphi_k(\tau)
&=&\frac1{\sqrt{2k}(2\pi)^{3/2}a(\tau)}
    e^{-ik\tau}\left(1-\frac i{k\tau}\right)\qquad\mbox{vacuum, matter},\\
&=&\frac 1{\sqrt{2k}(2\pi)^{3/2}a(\tau)}e^{-ik\tau}\qquad{\rm radiation}.
\end{eqnarray}
We used the fact that $\psi \approx {\rm const.}$ in the inflationary
phase (\ref{psiconst}), and $\psi = 0$ in the post inflationary
phase. We also assumed our universe approaches the conformal
vacuum mode function as $\tau \rightarrow -\infty$, that is
\begin{equation}
\varphi_{k,{\rm VD}}\rightarrow
\frac1{\sqrt{2k}(2\pi)^{3/2}a(\tau)}e^{-ik\tau}\qquad\mbox{for $\tau\rightarrow -\infty$}.
\end{equation}
Where the quantities with "VD" means the quantities evaluated at the vacuum dominated phase, namely at the inflationary phase.
The Bogoliubov coefficients relating operators at each phase are defined as :
\begin{eqnarray}
\hat{a}_{\rm RD}(\mbox{\boldmath$k$})&=&
c_1(k)\hat{a}_{\rm VD}(\mbox{\boldmath$k$})+
c_2^*(k)\hat{a}^{\dag}_{\rm VD}(-\mbox{\boldmath$k$}),\\
\hat{a}_{\rm MD}(\mbox{\boldmath$k$})&=&
c_3(k)\hat{a}_{\rm VD}(\mbox{\boldmath$k$})+
c_4^*(k)\hat{a}^{\dag}_{\rm VD}(-\mbox{\boldmath$k$}).
\end{eqnarray}
where "RD" and "MD" mean "Radiation Dominated" and "Matter Dominated", respectively. Matching the field and its first derivative at the first transition:
\begin{eqnarray}
\varphi_{k,\rm VD}(-\tau_2)&=&c_1\varphi_{k,\rm RD}(\tau_2)+c_2\varphi^*_{k,\rm RD}(\tau_2),\\
{\varphi}'_{k,\rm VD}(-\tau_2)&=&c_1{\varphi}'_{k,\rm RD}(\tau_2)+c_2{\varphi}^*{}'_{k,\rm RD}(\tau_2), 
\end{eqnarray}
we obtain
\begin{eqnarray}
c_1 &=& -\frac{H\tau_1}{(k\tau_0)^2}e^{2i k\tau_2}\left[1-2i k\tau_2-2(k\tau_2)^2\right],\\
c_2 &=& \frac{H\tau_1}{(k\tau_0)^2}.
\end{eqnarray}
We can see these coefficients satisfy the Bogoliubov relations
required for the orthonormality of mode functions $\varphi_k$\cite{BD} :
\begin{equation}
|c_1|^2 - |c_2|^2 =1.
\end{equation}
Since we are interested in GWs which are still well outside the horizon 
at the time of matter-radiation equality and will give the largest 
contribution to the CMBR anisotropy today, 
we may use simpler results up to second order
\begin{eqnarray}
c_1 &\approx& -\frac{H\tau_1}{(k\tau_0)^2},\\
c_2 &=& \frac{H\tau_1}{(k\tau_0)^2}.
\end{eqnarray}
At the next transition the continuity conditions become
\begin{eqnarray}
c_1\varphi_{k,\rm RD}(\frac{\tau_1}2)+c_2\varphi^*_{k,\rm RD}(\frac{\tau_1}2)&=&c_3\varphi_{k,\rm MD}(\tau_1)+c_4\varphi^*_{k,\rm MD}(\tau_1),\\
c_1{\varphi}'_{k,\rm RD}(\frac{\tau_1}2)+c_2{\varphi}^*{}'_{k,\rm RD}(\frac{\tau_1}2)&=&c_1{\varphi}'_{k,\rm MD}(\tau_1)+c_2{\varphi}^*{}'_{k,\rm MD}(\tau_1), 
\end{eqnarray}
Thus we obtain 
\begin{eqnarray}
c_3 &=& \frac{H/k}{2(k\tau_1)(k\tau_0)^2} \nonumber \\
\mbox{} & & \left[\left\{1-2ik\tau_1-2(k\tau_1)^2\right\}
            e^{ik\tau_1/2}\left\{1-2ik\tau_2-2(k\tau_2)^2\right\}
            e^{2ik\tau_2}-e^{3ik\tau_1/2}\right],\\
c_4 &=& \frac{H/k}{2(k\tau_1)(k\tau_0)^2} \nonumber \\
\mbox{} & & \left[e^{-3ik\tau_1/2}\left\{1-2ik\tau_2-2(k\tau_2)^2\right\}
            e^{2ik\tau_2}-\left\{1+2ik\tau_1-2(k\tau_1)^2\right\}
            e^{-ik\tau_1/2}\right],
\end{eqnarray}
with
\begin{equation}
|c_3|^2 - |c_4|^2 =1.
\end{equation}
Up to second order, we get simple solutions again as
\begin{eqnarray}
c_3&\approx&-\frac{3iH}{2k(k\tau_0)^2},\\
c_4&\approx& -\frac{3iH}{2k(k\tau_0)^2}.
\end{eqnarray}
Since we assume each transition is instantaneous, 
the Universe will remain in the de Sitter vacuum. 
Thus we may be able to calculate the quantum mechanical 
two-point correlation function in the de Sitter vacuum.
\begin{equation} 
\Delta^{\rm (QM)}\equiv \frac{k^3}{(2\pi)^2}\int d^3\mbox{\boldmath$x$}e^{i\mbox{\boldmath$k$}\cdot\mbox{\boldmath$x$}}\left<0|\varphi(\mbox{\boldmath$x$},\tau)\varphi(0,\tau)|0\right>,
\end{equation}
where 
\begin{equation}
\left. |0\right>\equiv \left. |\mbox{de Sitter vacuum}\right>.
\end{equation}
Note that this assumption makes us to use the Heisenberg picture of quantum fields in which the states of vacuum don't evolve with time but the operators do\cite{Allen}.
For waves re-entering the horizon at the matter dominated era, we have 
\begin{equation}
\Delta^{\rm (QM)}=\frac{H^2}{2(2\pi)^3}\left[\frac{3j_1(k\tau)}{k\tau}\right]^2,
\end{equation}
where $j_1(k\tau)$ is a spherical Bessel function of order one.
According to our normalization, this is calculated as 
\begin{equation}
\Delta^{\rm (QM)}_{\rm GW}{}_{,\lambda\lambda'}
=2\kappa_{\rm eff}^2 \delta_{\lambda\lambda'}\Delta^{\rm (QM)}
=\frac {\kappa^2H^2}{(2\pi)^3}
   \left[\frac{3j_1(k\tau)}{k\tau}\right]^2\delta_{\lambda\lambda'},
\end{equation}
where we used the following facts
\begin{eqnarray}
\epsilon_{\mu\nu}(\lambda)\epsilon^{\mu\nu}(\lambda')&=&\delta_{\lambda\lambda'},\\
\psi(\tau>\tau_2)&=&0.
\end{eqnarray}

We match this result with our classical ensemble of GWs. 
The classical two-point function may be calculated 
by writing the classical GWs as 
\begin{equation}
h_{\lambda}(\mbox{\boldmath$k$},\tau)
=A(k)\chi_{\lambda}(\mbox{\boldmath$k$})\left[\frac{3j_1(k\tau)}{k\tau}\right],
\end{equation}
where $\chi_{\lambda}$ is a random variable with statistical 
expectation value
\begin{equation}
\left<\chi_{\lambda}(\mbox{\boldmath$k$})\chi_{\lambda'}
(\mbox{\boldmath$k$}')\right>
=\frac1{k^3}\delta(\mbox{\boldmath$k$}-\mbox{\boldmath$k$}')
    \delta_{\lambda\lambda'}.
\end{equation}
Then the  classical two-point function is 
\begin{equation}
\Delta^{\rm (CL)}_{\rm GW}{}_{,\lambda\lambda'}
=A^2(k)\left[\frac{3j_1(k\tau)}{k\tau}\right]^2\delta_{\lambda\lambda'}.
\end{equation}
Comparing this with quantum two point function, we obtain
\begin{equation}
A^2(k)=\frac {\kappa^2H^2}{(2\pi)^3}=\frac8{3\pi}v,
\end{equation}
where 
\begin{equation}
H^2\equiv \frac{\kappa^2}3 m_pl^4 v.
\end{equation}
Since we know that $H^2 = \frac{\lambda\psi_i}{12\kappa^2\xi^2}$,
\begin{equation}\label{v}
v=\frac{\psi_i}{(16\pi)^2}\left(\frac{\lambda}{\xi^2}\right).
\end{equation}
In the end, we find that the existence of the nonminimal coupling
only affects the amplitude of spectrum via Hubble constant.

\section{Comparison With Observations}

We shall now make a comparison between the theoretical prediction 
derived in the previous section with the observation of CMBR anisotropy. 
Using the Sachs-Wolfe relation, one can calculate the power
spectrum of CMBR anisotropy generated by the long-wavelength GW \cite{White92}:
\begin{equation}
\left<a_\ell^2\right>\equiv\left<\sum_m |a_{\ell m}|^2\right>
=36\pi^2(2\ell+1)\frac{(\ell+2)!}{(\ell-2)!}\int_0^{2\pi/\tau_1} kdkA^2(k)|F_\ell(k)|^2,
\end{equation}
where the angle brackets denote averages over statistical ensemble of $a_{\ell}^2$, and
\begin{eqnarray}
F_\ell(k) 
& \equiv & \int^{\tau_0-\tau_1}_0 dr 
            \left[\frac d{d\left(k(\tau_0-r)\right)}
            \frac{j_1\left(k(\tau_0-r)\right)}{k(\tau_0-r)}\right] \nonumber \\ \mbox{} & & \times\left[\frac{j_{\ell-2}(kr)}{(2\ell-1)(2\ell+1)}
             +\frac{2j_{\ell}(kr)}{(2\ell-1)(2\ell+3)}
             +\frac{j_{\ell+2}(kr)}{(2\ell+1)(2\ell+3)}\right].
\end{eqnarray}
We shall use the following numerical value for $\left<a_2^2\right>$ 
evaluated by White\cite{White92}
\begin{equation}\label{a2}
\left<a_2^2\right> = 7.74v.
\end{equation}
On the other hand, COBE-DMR group expresses this quantity 
in terms of $Q_{\rm rms-PS}$ :
\begin{equation}\label{Qrms}
\left<a_2^2\right> \equiv 4\pi\left(\frac{Q_{\rm rms-PS}}{T_0}\right)^2.
\end{equation}
Note that since this quantity $Q_{\rm rms-PS}$ has already been handled statistically, we can make a direct comparison between this expression 
with the theoretical prediction $\left<a_{\ell}^2\right>$. 
Combining (\ref{v}), (\ref{a2}) and (\ref{Qrms}), we find
\begin{equation}
\frac{\lambda}{\xi^2} = \frac{(4\pi)^3}{38.7}\left(\frac{80}{\psi_i}\right)\left(\frac{Q_{\rm rms-PS}}{T_0}\right)^2.
\end{equation}
According to COBE 4yr results\cite{Fixsen,Bennett}, 
\begin{eqnarray}
T_0 &=& 2.728\pm 0.004 \ {\rm K}\label{cobe1},\\
Q_{\rm rms-PS} &=& 18\pm 1.4\ {\rm \mu K}\label{cobe2},
\end{eqnarray}
for $n=1$ Harrison-Zel'dovich spectrum, thus we obtain the following 
constraint on the theory.
\begin{equation}\label{final}
\frac{\lambda}{\xi^2} \approx 
2.3 \times 10^{-9}\left(\frac{80}{\psi_i}\right),
\end{equation}
where we do not take the errors of the observations (\ref{cobe1}) and
(\ref{cobe2}) into account, because our result (\ref{final})
mainly depends on the initial value of $\psi_i$. 
Furthermore, we have not considered the contribution from 
the scalar density perturbation which will be produced during early 
stage of the inflationary phase. Although a
large fraction of the quadrupole anisotropy seems to be due to GW\cite{KW}, 
we have to take that contribution into account for more precise discussions. Note that since the uncertainties just mentioned lead to higher estimates of $\lambda/\xi^2$, we may 
expect that the ``real value'' of $\lambda/\xi^2$ is slightly lower than 
the obtained result (\ref{final}).

Makino and Sasaki\cite{MS91} derived the perturbation of spatial curvature ${\cal R}_m$ on the hypersurfaces orthogonal to the matter rest frame :
\begin{equation}  
{\cal R}_{m} = \frac{N(t_k)}{\pi}\sqrt{\frac{\lambda}{3\xi(1+6\xi)}},
\end{equation}
where $N(t_k)$ is a e-fold scale in which one is interested.
${\cal R}_m$ is related to the gauge-invariant density perturbation as
\begin{equation}
\frac{\delta\rho}{\rho}
=\frac{2(w+1)}{3w+5}\left(\frac{k}{Ha}\right)^2{\cal R}_m,
\end{equation}
where $w\equiv p/\rho$. Using our result (\ref{final}), the density
perturbation on the horizon scale ($k=aH$, $N = 70$) at a matter-dominated phase ($w=0$) becomes
\begin{equation}
\left(\frac{\delta\rho}{\rho}\right)_{\rm HOR}\approx 2.1\sqrt{\frac{\lambda}{\xi^2}} \approx 1.0\times 10^{-4}\sqrt{\frac{80}{\psi_i}}.
\end{equation}
The density perturbation is also related to the rms temperature
fluctuation of CMBR via Sachs-Wolfe effect as
\begin{equation}
\left(\frac{\delta T}{T}\right)_{\rm rms}=-\frac13 \Phi,
\end{equation}
where $\Phi$ is the Newton potential, i.e. $g_{00}=1-2\Phi$. 
On the other hand, $\Phi$ is determined by the Poisson equation,
\begin{equation}
k^2
=-4\pi G\rho\left(\frac{\delta\rho}{\rho}\right)
=\frac32(aH)^2\left(\frac{\delta\rho}{\rho}\right).
\end{equation}
Since our interest is the density perturbation on present horizon
scale, i.e. $k=aH$, we have 
\begin{equation}
\left(\frac{\delta T}{T}\right)_{\rm rms}
=\frac12\left(\frac{\delta\rho}{\rho}\right)_{\rm HOR}
\approx \sqrt{\frac{\lambda}{\xi^2}}.
\end{equation}
 From the COBE-DMR result $(\delta T/T)_{\rm rms}=1.1\times
10^{-5}$\cite{Bennett}, we obtain the {\sl initial condition of our
universe} as
\begin{equation}
\psi_i \approx 1.6\times 10^3.
\end{equation}
We {\sl must} notice again, however, we have never paid attention to
the ratio of tensor contribution to scalar one. Since smaller
tensor contribution leads to  smaller value of $\psi_i$ by several factors, 
our discussion is reliable up to order of magnitude.

Finally, let us consider the model of $\lambda=10^{-2}$, which
would be a reasonable value for some of particle physics model. 
Then the nonminimal coupling constant and the initial value of the 
inflaton field are constrained as
\begin{eqnarray}
\xi&\approx&10^4,\\
\frac{\phi_i}{m_pl}&\approx&10^{-1},
\end{eqnarray}
As Makino and Sasaki commented\cite{MS91}, the relaxation of
fine-tuning for $\lambda$ is nothing but a restatement of the issue 
of a large values of $\xi$. According to our results, furthermore, the
initial inflaton field has to be smaller by one order of magnitude 
than that of chaotic scenario in framework of GR. Note that the
smaller contribution of tensor perturbation leads to the smaller
$\psi_i$ and then smaller $\phi_i/m_pl$.

\section{Conclusions}

We calculated the spectrum of GWs generated in the chaotic inflationary 
scenario with a nonminimally coupled inflaton under the condition where $\psi\equiv \kappa^2\xi\phi^2\gg 1$, 
and their contributions to the CMBR anisotropy via Sachs-Wolfe effect. 
Comparing the COBE-DMR 4yr observations with the predicted
power spectrum of CMBR, we find the initial value $\psi_i \approx
1.6\times 10^3$. Our result decreases a number of undetermined
parameters in the theory, i.e. $\psi_i$, and indicates that if we choose
$\lambda=10^{-2}$, $\xi\approx 10^4$ and $\phi_i/m_pl\approx
10^{-1}$ are required. 
In the chaotic inflationary scenario with minimally coupled inflaton, an argument based on the comparison between the energy content of the inflaton field and planck  energy density places the upper bound of 
the initial value of the field. 
It would be hoped in the nonminimally coupled case that the same sort 
of argument will place an upper bound on the initial value for $\psi_i$. 
However it is the nonminimality that prevent us to have such a simple 
argument. In this context Hochberg and Kephart made some discussion 
on the energy density of a nonminimally coupled scalar field\cite{HK}, 
but their result seems to be not available in the present context 
because of the existence of self-coupling. It should be mentioned that 
the condition for Hubble constant to be smaller than the inverse of the planck 
time is easily satisfied in the above choice of parameters. In this sense 
Fakir and Unruh  scenario is not excluded as a good candidate of the inflationary scenario.

Finally we mention the chaotic inflationary scenario by induced
gravity\cite{FU90b,AZT}. 
The above analysis may be applied exactly same way and one has the same 
constraint of parameters when $\psi \gg 1$. However this case has also another constraint 
on $\xi$ coming from solar system experiment on the omega parameter of Brans-Dicke theory since the theory is translated into Brans-Dick type. 
These two constraints are incompatible each other so that the chaotic scenario 
by induced gravity with the {\sl strong} coupling seems not to work.

\section*{Acknowledgments}

We would like to thank N. Sugiyama for very important comments and
suggestions. We would also like to thank M. Takada for useful discussions.

\end{document}